# Access to care improves EHR reliability and clinical risk prediction model performance

Anna Zink, Hongzhou Luan, Irene Y. Chen

**Abstract**
Disparities in access to healthcare have been well-documented in the United States, but their effects on electronic health record (EHR) data reliability and resulting clinical models are poorly understood. Using an All of Us dataset of 134,513 participants, we investigate the effects of access to care on the medical machine learning pipeline, including medical condition rates, data quality, outcome label accuracy, and prediction performance. Our findings reveal that patients with cost constrained or delayed care have worse EHR reliability as measured by patient self-reported conditions for 78% of examined medical conditions. We demonstrate in a prediction task of Type II diabetes incidence that clinical risk predictive performance can be worse for patients without standard care, with balanced accuracy gaps of 3.6 and sensitivity gaps of 9.4 percentage points for those with cost-constrained or delayed care. We evaluate solutions to mitigate these disparities and find that including patient self-reported conditions improved performance for patients with lower access to care, with 11.2 percentage points higher sensitivity, effectively decreasing the performance gap between standard versus delayed or cost-constrained care. These findings provide the first large-scale evidence that healthcare access systematically affects both data reliability and clinical prediction performance. By revealing how access barriers propagate through the medical machine learning pipeline, our work suggests that improving model equity requires addressing both data collection biases and algorithmic limitations. More broadly, this analysis provides an empirical foundation for developing clinical prediction systems that work effectively for all patients, regardless of their access to care.

**Introduction**
The ability to access healthcare fundamentally shapes health outcomes, yet persistent social and economic barriers keep essential care out of reach for many Americans, especially people of color and other marginalized groups.[1,2] As healthcare systems increasingly leverage large datasets to deploy predictive models to guide clinical decisions[3–6], this disparity creates a concerning paradox: the patients most affected by access barriers may be systematically misrepresented in the very data used to develop these models. Despite extensive documentation of healthcare access disparities, their effects on electronic health records and resulting clinical predictions remain largely unknown.

Previous studies suggest that limited access reduces model performance but rely on indirect proxies like race or insurance status rather than direct measures of access.[7–11] This knowledge gap is particularly concerning as healthcare systems increasingly base clinical decisions on machine learning models trained using electronic health record data. Understanding how access shapes both data collection and algorithmic performance requires datasets that directly measure healthcare access and link it to clinical records.

In our study, we analyze survey data and electronic health record data from 134,513 participants in the NIH All of Us program to quantify these relationships.[12] Developed by the National Institute of Health, the All of Us program is one of the largest biomedical data resources in the United States with the explicit goal of creating a diverse dataset. Because the data include self-reported information on access to care linked with EHR data, it provides a unique opportunity to study the implications of access to care on clinical risk prediction.

We first quantify how access to care affects the reliability of EHR data by comparing self-reported condition data with conditions recorded in the EHR. Then we evaluate the extent to which clinical machine learning methods may create biased predictions for patients with lower access to care. We focus on one clinical task: 2-year Type 2 diabetes incidence. This task was selected because Type 2 diabetes is one of the most common chronic conditions in the United States today and early identifying and treatment of diabetes could improve health outcomes for many patients.[13] The area of disease incidence is also widely recognized as an area in which machine learning and predictive analytics have the potential to improve clinical targeting.[14–17]

*Sample*
Our sample included 134,513 All of Us participants for which we had self-reported data on healthcare access and linked EHR data (Figure 1b). Using the healthcare access survey, we identified participants that had delayed care or hadn't obtained care due to affordability concerns. More than half of the participants indicated that they had trouble accessing care, either reporting affordability problems (42.9%) and/or that they had delayed care (34.6%) in the previous year. Cited reasons for delaying care included having to pay out of pocket, nervousness, and inability to take time off from work. Cited affordability issues included having to take a lower cost prescription to save money, having to pay out of pocket, and nervousness about affording care (Figure S1).

We classified participants into three groups based on their responses to the healthcare access survey. We defined a participant as having "standard care" to healthcare services if they indicated that they did not have issues with affording or delaying care. Participants that indicated affordability concerns were categorized as having "cost-constrained care" and participants indicating that they had delayed care were categorized as having "delayed care". For this work, we refer to patients with either cost-constrained care or delayed case as having "low access" to care. White participants were much more likely to report standard levels of care compared to Black and Hispanic participants (Table S1).

In Table 1, we compare participants with different levels of care. Participants with cost-constrained and or delayed care were more likely to be younger, female, Black, and Hispanic. We computed age-adjusted prevalence rates of the 15 most-common self-reported conditions by access group (Table S2). Participants with standard care were less likely to report 14 out of 15 considered conditions. For example, participants with standard care had lower age-adjusted rates of hypertension, depression, anxiety, and asthma. All differences were significant at the $p<0.05$ level, based on a two-sided t-test.

*Patients with lower access to care have lower EHR reliability*
Data quality issues in the EHR are a well-explored topic[18], with many studies documenting the effect of EHR quality on research results.[19,20] Prior research has found high rates of discordance between survey data and EHR data on data elements including health conditions and demographics.[21,22] A recent study on primary care documentation also found that patient-initiated issues were often omitted from notes.[23]

We examined reported rates of 37 health conditions in the EHR versus self-report (Figure S2). There was significant variation across conditions in terms of likelihood of appearance in the EHR, self-report or both. This was not necessarily surprising. Patients may be less likely to report certain conditions if they are unfamiliar with the medical term (e.g., aortic aneurysm was more likely to be reported in the EHR and not by the participant) or uncomfortable sharing certain medical information due to social stigma[24,25] (e.g., drug use disorder was more commonly reported in the EHR and not by the participant). On the other hand, the EHR may be missing certain conditions that were self-reported if the patient had not been in for a medical visit and/or the condition had not been recorded during a visit.

To measure EHR reliability, we focused on the rate of self-reported conditions that were not reported in the EHR. Participants with cost-constrained or delayed care had statistically significant higher rates of EHR missingness for 29 out of 37 conditions compared to the standard care group (Figure 2), computed using a chi-squared test with a Benjamini-Hochberg false discovery correction of $\alpha=0.05$.[26] For example, among common conditions such as depression and hypertension, which were reported by more than 25% of participants, we observed significantly higher rates of missingness for low versus high access groups (depression: 64% versus 60%, hypertension: 47% versus 41%).

On average, low access participants were missing 2.0 diagnoses in the EHR versus 1.7 for those with high access (p-value: <0.001). When we ran a regression predicting the presence of an EHR condition given the presence of the self-reported condition and access variable, controlling for conditions, we found that access to care was negatively correlated with the presence of an EHR condition by 1% (p-value <0.001). If we only looked in the EHR in the prior year (a common lookback window in prediction models), this amount doubled to 2%.

*Algorithmic predictive performance is lower for patients without access to care*
When EHR data is used for predictive algorithms, differences in medical condition reliability indicate that patients with lower access to care may have worse quality feature data or outcome labels. Lower quality data could degrade the predictive quality for those with lower access to care, decreasing the ability of the model to identify low access patients at risk of an event.

We assessed the impact of access to care on predictive performance on a prediction model of 2-year Type 2 diabetes incidence. The prediction task was set up from the perspective of the health system, using only data available in the EHR. The prediction sample included 52,046 participants that did not have a record of type II diabetes in the EHR at the time they completed their baseline survey for All of Us. The rate of Type 2 diabetes incidence was 0.6%. We noted the same pattern of demographic differences between standard care, cost-constrained care, and delayed care access groups; namely, that cost-constrained and delayed care access groups were more likely to be younger, female, Black, and Hispanic (Table S4). We used a 2-year lookback window to construct the features, which included clinical conditions, clinical measures (i.e., labs and vitals), number of procedures, number of medications, and demographic information (age, gender, race, and ethnicity).

The logistic regression model with an L1 regularization (or LASSO) outperformed the random forests with an AUC of 0.754 (95% CI: 0.726-0.783). In Figure 3, we plot performance results from the LASSO algorithm, stratified by access to care levels. Although the AUC was lower for the cost-constrained and delayed care group, we found no significant difference between the three groups. However, when we looked at the classification metrics, we saw a significant drop in balanced accuracy for the cost-constrained and delayed care group compared to the standard care group. This drop was due to a large reduction in sensitivity, and only modest increases in specificity. For example, the model was only able to classify 57.0% (95% CI: 56.3-57.7) of new diabetes cases for the delayed care group and 58.2% (95% CI: 57.6 - 58.9) for the cost-constrained group relative to 66.4% (95% CI: 65.9-67.0) for the standard care group. Lower sensitivity may lead to less comparative clinical interventions for cost-constrained and delayed care groups, because the model will identify less of these patients as high-risk for diabetes incidence.

*Potential Solutions*
We evaluated two strategies to improve model performance for patients with cost-constrained and delayed care: (1) incorporating an indicator for standard care as a feature in the clinical prediction model and (2) including self-reported condition data as features in the model (Figure 4).

Incorporating standard care as a feature could improve the predictive performance for those with cost-constrained or delayed care if there was differential risk of type 2 diabetes incidence for those with low access to care which was not captured by the features in the model.[27] However, when we included it as a feature, we did not find any difference in AUC. Although, we did see a small, but significant increase in balanced accuracy and sensitivity for the delayed care group.

Our second approach aimed to improve the quality of the feature data used in the prediction model. In this example, including self-reported conditions as features could improve the predictive performance by providing information to the model that was not available in the EHR features. When we included all self-reported conditions in the diabetes prediction model, we observed an increase in AUC for the cost-constrained and delayed care groups, but this difference was not statistically significant. However, balanced accuracy and sensitivity increased significantly for both groups. For example, sensitivity in the cost-constrained group increased by more than 11.2 percentage points (from 58.2% to 69.4%, p-value <0.005). Furthermore, balanced accuracy and sensitivity gaps between standardized care and cost-constrained and delayed care were greatly diminished or disappeared entirely.

**Discussion**
The relationship between access to care and clinical prediction has been relatively understudied to-date, despite the prevalence of access issues in the U.S.[1,2] paired with an increasing reliance on prediction models to aid in clinical prediction models.[3] Here, we use novel data from the All of Us platform to study how access to care may affect EHR reliability and clinical prediction performance. We found consistently lower EHR reliability across nearly thirty self-reported health conditions for those with cost-constrained or delayed care, which confirms concerns related to the quality of EHR data for those with lower access to healthcare services.

This appears to have consequences for clinical prediction. In a clinical prediction task of type II diabetes incidence, we observed lower predictive performance for those with cost-constrained or delayed care as measured through balanced accuracy and sensitivity. Further analysis suggested that this performance gap was partly driven by missing and noisy data: when we incorporated self-reported condition data into the type 2 diabetes model, we significantly improved the predictive performance for those with cost-constrained or delayed care.

This work relates to a growing body of work on the measurement and mitigation of potential sources of algorithmic bias in clinical prediction models.[28] While previous work has focused on bias with respect to protected classes such as those defined by race, ethnicity, or gender, here we focus on access as a potential mechanisms for reduced performance across groups. To date there has been a lack of empirical research quantifying the relationship between healthcare access and clinical prediction, in part due to data limitations in EHR and health claims data which do not record measures of healthcare access. Our findings provide quantitative evidence to support the concerns around reduced prediction performance for those with lower access to care and suggest that collecting better, patient-centered data could help alleviate this disparity. This coincides with a growing movement of incorporating patient-reported outcomes into prediction models.[29,30]

There are several limitations of this work. First, we relied on self-reported data to validate the reliability of electronic health records (EHR). Self-reported data have their own inherent limitations and may not always be accurate.[31] Second, EHR data were limited to clinical sites participating in the All of Us program. Third, we were unable to validate the accuracy of the outcome labels used in the predictive task. As noted earlier, these labels may be imprecise especially for individuals with lower access to care, leading to noisier measurements. Finally, the All of Us program is a convenient sample and not a representative sample of patients in the U.S. This study raises critical concerns about the limitations of clinical risk prediction models for individuals with lower access to care. Our findings suggest that incorporating new data modalities, such as patient-reported health outcomes, could help mitigate this issue. However, this type of information is typically unavailable in most clinical settings. Future research should explore methodological solutions that do not depend on collecting new data. Possible approaches include imputing missing data for those with low access to care and leveraging proxies for healthcare access within existing EHR data, such as clinical notes, social determinants of health (SDOH) data, and visit history.[32,33] Additionally, further investigation is needed to understand the downstream clinical consequences of reduced model performance for individuals with limited healthcare access. This work underscores the limitations of relying on EHR data for risk prediction and highlights the value of integrating multi-modal, patient-centered data sources to quantify, understand, and address health disparities associated with healthcare access.

**Material and Methods**

*Data & Sample*

Our data come from the All of Us database.[12] The All of Us database is a convenient sample of over a million people across 50 states. Data include survey data and (contingent on participant agreement) linked EHR data from participating provider organizations in the All of Us program. Several surveys are administered to participants upon their enrollment in the All of Us program. Surveys include questions on participant demographics ("The Basics" survey), health history ("Personal and Family Health History" survey), and health care access ("Health Care Access & Utilization" survey). EHR data are standardized using the Observational Medical Outcomes

Partnership (OMOP) Common Data Model infrastructure and are current through July 1, 2022. Our analysis sample includes all participants with linked EHR data that answered the "Health Access & Utilization" survey.

We are interested in understanding how access impacts clinical prediction. Timely access to care may be delayed for both financial and non-financial reasons. We use the "Health Access & Utilization" survey to define health care access for participants. The survey includes questions about affordability and reasons for delaying care within the last 12 months, including transportation issues, childcare, and inability to take off work. We defined a participant as having "standard care" to healthcare services if they indicated that they did not have issues with affording or delaying care. We define a participant as having "cost-constrained care" if they indicated that they had issues affording care, and "delayed care" if they had delayed care.

We define participant age, gender, race, and ethnicity using data collected from "The Basics" survey. Participant age was calculated using the date of birth reported by the participant. Participant gender was measured based on the participants' description of their gender identity. We defined a participant as Black if they responded "Black, African American, or African" to the question "Which categories describe you?". We defined a participant as Hispanic if they responded "Hispanic, Latino, or Spanish" to the same question. Participants could indicate multiple categories.

*Comparing Self-Reported Conditions to EHR Conditions*
To test rates of concordance by access to care, we compared self-reported conditions in the All of Us "Personal and Family Health History" survey to health conditions documented in the EHR. EHR data are organized into concepts under the OMOP format. We identified EHR conditions by searching for all condition concepts that matched the given keyword. Please see Table S5 for a list of SNOMED codes used to identify conditions in the EHR and the relevant survey question used to identify the self-reported condition.

We quantified the percent of participants with a self-reported condition that had no record of the same condition in the EHR (or "missing EHR diagnosis rate") for 37 self-reported conditions. There are two margins of discordance between these values. First, a participant might say they have a condition, but no record exists in the EHR. Second, a participant might say they do not have a condition, but there is a record of it in the EHR. We focused on the former: conditions that the participant says they have, assuming that the patient is the source of truth, and that any missing EHR record of a condition is an error.

We quantify missing EHR diagnosis rates for each of the 37 self-reported conditions by calculating the percent of participants with a self-reported condition reported in the "Personal and Family Health History" survey that had no record of the same condition in the EHR. Our main calculation considers whether a condition is ever present in the EHR, but we also assess how rates of missingness change using a 1-year lookback window, since observation window can vary across many prediction models.[34]

To summarize the correlation between access and EHR reliability, we test whether the mean number of missing conditions for those with low versus high access to care differs. Finally, we run a linear regression model to assess how low access modifies the relationship between a condition being reported in the EHR and by the participant, controlling for condition (i.e. $EHR = \alpha\, Self + \beta\, Low\, Access + \gamma\, Condition + \epsilon$) and report the coefficient on $\beta$.

*Implications for Clinical Prediction Models*

The diabetes sample included any participant with no EHR-record of Type II Diabetes as of the time in which the participant completed the "Personal and Family Health History" survey. If the participant did not complete this survey, then we use the date that the participant filled out the "Health Care Access and Utilization Survey" survey.[9] We refer to this date as the index date.

The outcome is an indicator for whether someone had a record of Type II diabetes in the two years following the index date (1 = Yes, 0 = No). Features are constructed from the EHR using a two-year lookback window from the index date. This included indicators for clinical conditions, clinical measures (obesity, diabetes, fever, hypocalcemia, hypercalcemia, hypochloremia, hyperchloremia, creatine, high blood pressure, tachycardia, anemia, high hemoglobin, hyperkalemia, hypokalemia, tachypneic, bradypnea, hypernatremia, hyponatremia, low urea, and high urea, see Table S2 for lab values and ranges for each measure), number of procedures, number of medications, and demographics (age, gender, race/ethnicity).

For the prediction task, logistic regression and random forests were trained with 5-fold cross validation. Logistic regression models were developed with a regularization parameter search from 1e-4 to 1e4 in ten logarithmically spaced intervals and across L1 and L2 regularization. Random forest models were developed with a regularization parameter search over 5, 50, 500, and 1000 decision trees. For classification, we selected the cutoff point that maximized Youden's J statistic. Model performance (AUC and balanced accuracy) was assessed overall and by access levels. For uncertainty quantification, we compute the numerical 95% confidence intervals as derived from the Central Limit Theorem.

**Acknowledgements**
We gratefully acknowledge *All of Us* participants for their contributions, without whom this research would not have been possible. We also thank the National Institutes of Health's *All of Us* Research Program for making available the participant data examined in this study. This study used data from the *All of Us* Research Program's Controlled Tier Dataset v7, available to authorized users on the Researcher Workbench. IC and AZ are funded by a Google Research Scholar award for this work. IC receives additional support from an Apple Machine Learning Faculty Research Award. AZ receives additional support from Chicago Booth Center for Applied AI.

**References**

1. Call, K. T. *et al.* Barriers to care in an ethnically diverse publicly insured population: is health care reform enough? *Med Care* **52**, 720–727 (2014).

2. Caraballo, C. *et al.* Trends in Racial and Ethnic Disparities in Barriers to Timely Medical Care Among Adults in the US, 1999 to 2018. *JAMA Health Forum* **3**, e223856 (2022).

3. Davenport, T. & Kalakota, R. The Potential for Artificial Intelligence in Healthcare. *Future Healthc J* **6**, 94–98 (2019).

4. Gervasi, S. S. *et al.* The Potential For Bias In Machine Learning And Opportunities For Health Insurers To Address It. *Health Affairs* **41**, 212–218 (2022).

5. Ghassemi, M. *et al.* A Review of Challenges and Opportunities in Machine Learning for Health. *AMIA Jt Summits Transl Sci Proc* **2020**, 191–200 (2020).

6. Shen, J. H., Raji, I. D. & Chen, I. Y. The Data Addition Dilemma. Preprint at https://doi.org/10.48550/arXiv.2408.04154 (2024).

7. Seyyed-Kalantari, L., Zhang, H., McDermott, M. B. A., Chen, I. Y. & Ghassemi, M. Underdiagnosis bias of artificial intelligence algorithms applied to chest radiographs in under-served patient populations. *Nat Med* **27**, 2176–2182 (2021).

8. Obermeyer, Z., Powers, B., Vogeli, C. & Mullainathan, S. Dissecting racial bias in an algorithm used to manage the health of populations. *Science* **366**, 447–453 (2019).

9. Chen, I. Y., Joshi, S. & Ghassemi, M. Treating health disparities with artificial intelligence. *Nat Med* **26**, 16–17 (2020).

10. Chen, I. Y., Szolovits, P. & Ghassemi, M. Can AI Help Reduce Disparities in General Medical and Mental Health Care? *AMA Journal of Ethics* **21**, 167–179 (2019).

11. Chen, I. Y. *et al.* Ethical Machine Learning in Healthcare. *Annual Review of Biomedical Data Science* **4**, null (2021).

12. Murray, J. The 'All of Us' Research Program. *N Engl J Med* **381**, 1884 (2019).

13. Pratley, R. E. The Early Treatment of Type 2 Diabetes. *The American Journal of Medicine* **126**, S2–S9 (2013).

14. Razavian, N. *et al.* Population-Level Prediction of Type 2 Diabetes From Claims Data and Analysis of Risk Factors. *Big Data* **3**, 277–287 (2015).

15. Park, J. H. *et al.* Machine learning prediction of incidence of Alzheimer's disease using large-scale administrative health data. *npj Digit. Med.* **3**, 1–7 (2020).

16. Zhang, Y. *et al.* Applying Artificial Intelligence Methods for the Estimation of Disease Incidence: The Utility of Language Models. *Front. Digit. Health* **2**, (2020).

17. Delpino, F. M. *et al.* Machine learning for predicting chronic diseases: a systematic review. *Public Health* **205**, 14–25 (2022).

18. Sauer, C. M. *et al.* Leveraging electronic health records for data science: common pitfalls and how to avoid them. *Lancet Digit Health* **4**, e893–e898 (2022).

19. Bower, J. K., Patel, S., Rudy, J. E. & Felix, A. S. Addressing Bias in Electronic Health Record-Based Surveillance of Cardiovascular Disease Risk: Finding the Signal Through the Noise. *Curr Epidemiol Rep* **4**, 346–352 (2017).

20. Beaulieu-Jones, B. K. *et al.* Machine learning for patient risk stratification: standing on, or looking over, the shoulders of clinicians? *npj Digit. Med.* **4**, 1–6 (2021).

21. Wagaw, F. Linking Data From Health Surveys and Electronic Health Records: A Demonstration Project in Two Chicago Health Center Clinics. *Prev. Chronic Dis.* **15**, (2018).

22. O'Brien, E. C. *et al.* Concordance Between Patient-Reported Health Data and Electronic Health Data in the ADAPTABLE Trial. *JAMA Cardiol* **7**, 1235–1243 (2022).

23. Weiner, M. *et al.* Accuracy, thoroughness, and quality of outpatient primary care documentation in the U.S. Department of Veterans Affairs. *BMC Primary Care* **25**, 262 (2024).

24. Magura, S. & Kang, S. Y. Validity of self-reported drug use in high risk populations: a meta-analytical review. *Subst Use Misuse* **31**, 1131–1153 (1996).

25. Darke, S. Self-report among injecting drug users: A review. *Drug and Alcohol Dependence* **51**, 253–263 (1998).

26. Benjamini, Y. & Yekutieli, D. The control of the false discovery rate in multiple testing under dependency. *The Annals of Statistics* **29**, 1165–1188 (2001).

27. Zink, A., Obermeyer, Z. & Pierson, E. Race adjustments in clinical algorithms can help correct for racial disparities in data quality. *Proc Natl Acad Sci U S A* **121**, e2402267121 (2024).

28. Huang, J., Galal, G., Etemadi, M. & Vaidyanathan, M. Evaluation and Mitigation of Racial Bias in Clinical Machine Learning Models: Scoping Review. *JMIR Med Inform* **10**, e36388 (2022).

29. Rivera, S. C. *et al.* Embedding patient-reported outcomes at the heart of artificial intelligence health-care technologies. *The Lancet Digital Health* **5**, e168–e173 (2023).

30. Paudel, R., Dias, S., Wade, C. G., Cronin, C. & Hassett, M. J. Use of Patient-Reported Outcomes in Risk Prediction Model Development to Support Cancer Care Delivery: A Scoping Review. *JCO Clin Cancer Inform* e2400145 (2024) doi:10.1200/CCI-24-00145.

31. Martin, L. M., Leff, M., Calonge, N., Garrett, C. & Nelson, D. E. Validation of self-reported chronic conditions and health services in a managed care population. *Am J Prev Med* **18**, 215–218 (2000).

32. Getzen, E., Ungar, L., Mowery, D., Jiang, X. & Long, Q. Mining for equitable health: Assessing the impact of missing data in electronic health records. *Journal of Biomedical Informatics* **139**, 104269 (2023).

33. Zhang, H., Clark, A. S. & Hubbard, R. A. A Quantitative Bias Analysis Approach to Informative Presence Bias in Electronic Health Records. *Epidemiology* **35**, 349–358 (2024).

34. Lauritsen, S. M. *et al.* The Framing of machine learning risk prediction models illustrated by evaluation of sepsis in general wards. *NPJ Digit Med* **4**, 158 (2021).

**Table 1. Study Sample Characteristics**

|  | Overall (N=134,513) | High Access (N=62,454) | Affordability (N=57,654) | Delayed Care (N=46,582) |
|---|---|---|---|---|
| **Access Measures (%)** | | | | |
|   Affordability | 42.9 | 0 | 100.0 | 69.1 |
|   Delayed Care | 34.6 | 0 | 55.8 | 100.0 |
| **(Average) Age** | 54.8 | 59.4 | 51.2 | 47.3 |
| **Gender (%)** | | | | |
|   Male | 33.5 | 39.4 | 28.1 | 25.7 |
|   Female | 63.7 | 58.2 | 68.6 | 70.6 |
|   Other | 2.8 | 2.4 | 3.3 | 3.7 |
| **Race (%)** | | | | |
|   White | 70.3 | 75.4 | 65.1 | 64.6 |
|   Black | 10.4 | 08.8 | 12.1 | 12.3 |
|   Other | 19.3 | 15.7 | 22.8 | 23.0 |
| **Ethnicity (%)** | | | | |
|   Hispanic or Latino | 12.1 | 8.44 | 15.8 | 15.8 |

Includes All of Us participants that answered the Healthcare and Access survey and had linked EHR data.

**Figure 1: Study overview and data cohort.**

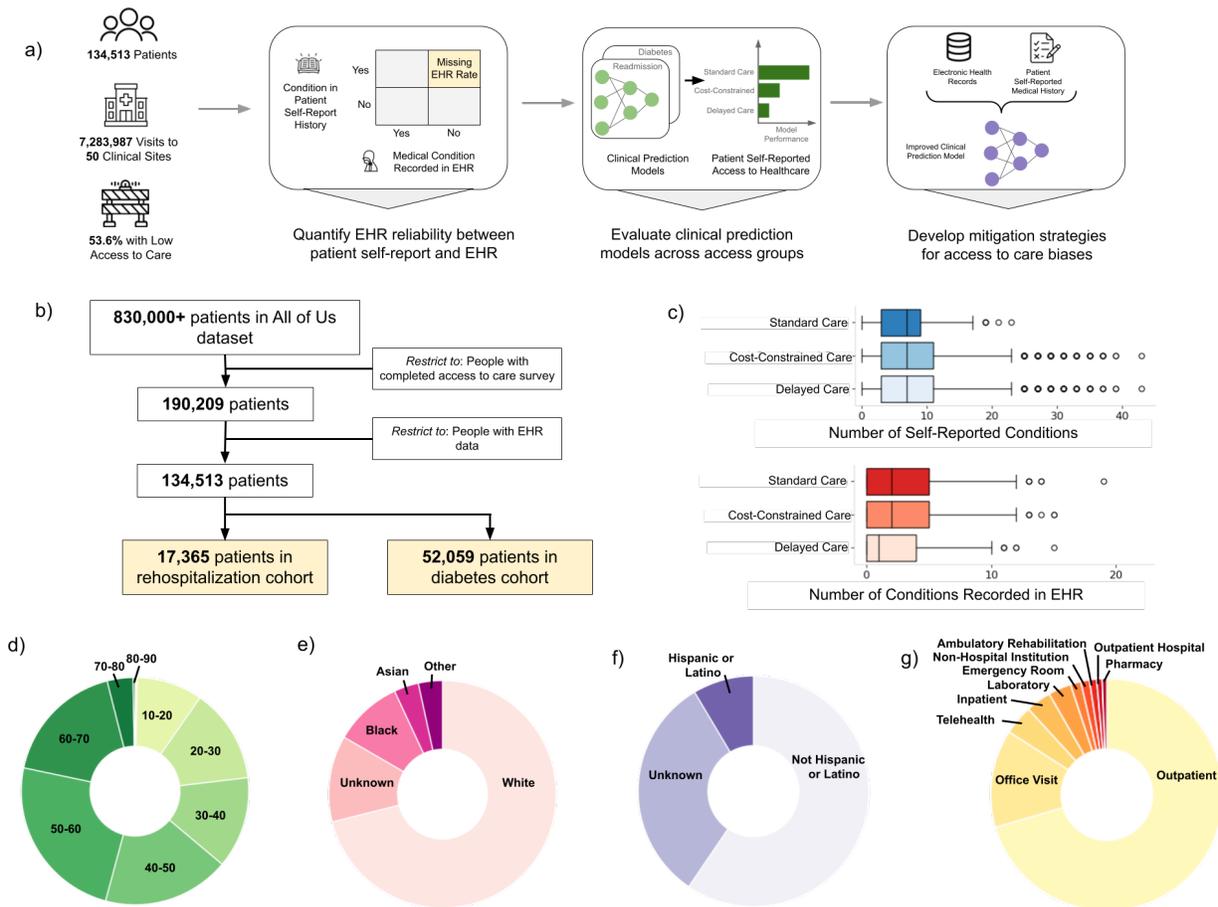

Overview of paper approach to determine EHR reliability and clinical model performance bias due to differences in access to healthcare. (a) Using the All of Us dataset, we quantify EHR reliability by comparing patient self-reported medical history with electronic health record (EHR) medical conditions. We then assess the impact of differences in EHR reliability by evaluating trained clinical prediction models across access to care patient groups. Lastly, we develop mitigation strategies for these biases due to access to care using both EHR data and patient self-reported medical history. (b) Patient cohort creation flowchart beginning with the full All of Us patient dataset and excluding patients according to access to care survey completion and EHR data availability. (c) Distribution of self-reported diagnoses and EHR conditions per individual across access to care patient groups in our patient cohort. (d-f) Number of patients in each (d) age category, (f) race category, and (g) ethnicity category. (g) Number of EHR records in each visit type category.

**Figure 2: Missing EHR Diagnosis Rate in Low and High Access by Condition**

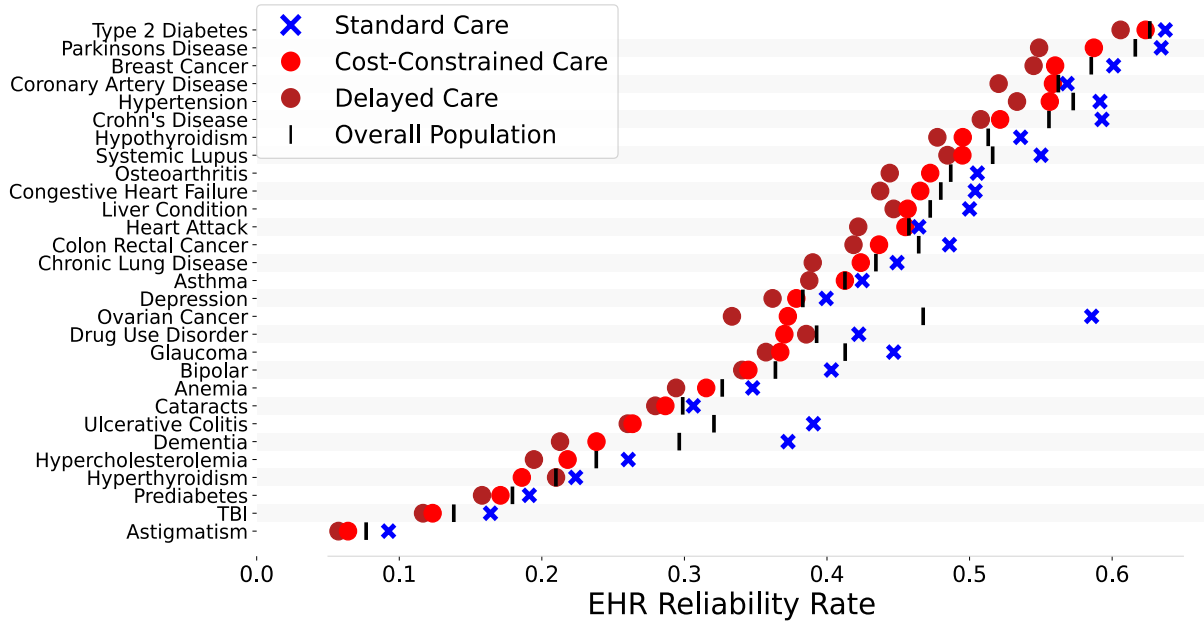

We quantify EHR reliability by comparing patient self-reported medical history with electronic health record (EHR) medical conditions. We define the missing EHR diagnosis rate as the share of participants with a self-reported condition who did not have the conditions documented in the EHR. In the figure we show all conditions with a statistically significant (p-value <0.05 with t-test comparisons) difference in missing EHR diagnosis rates between the low access and high access group.

**Figure 3: Model Performances Stratified by Access to Care for Diabetes Task**

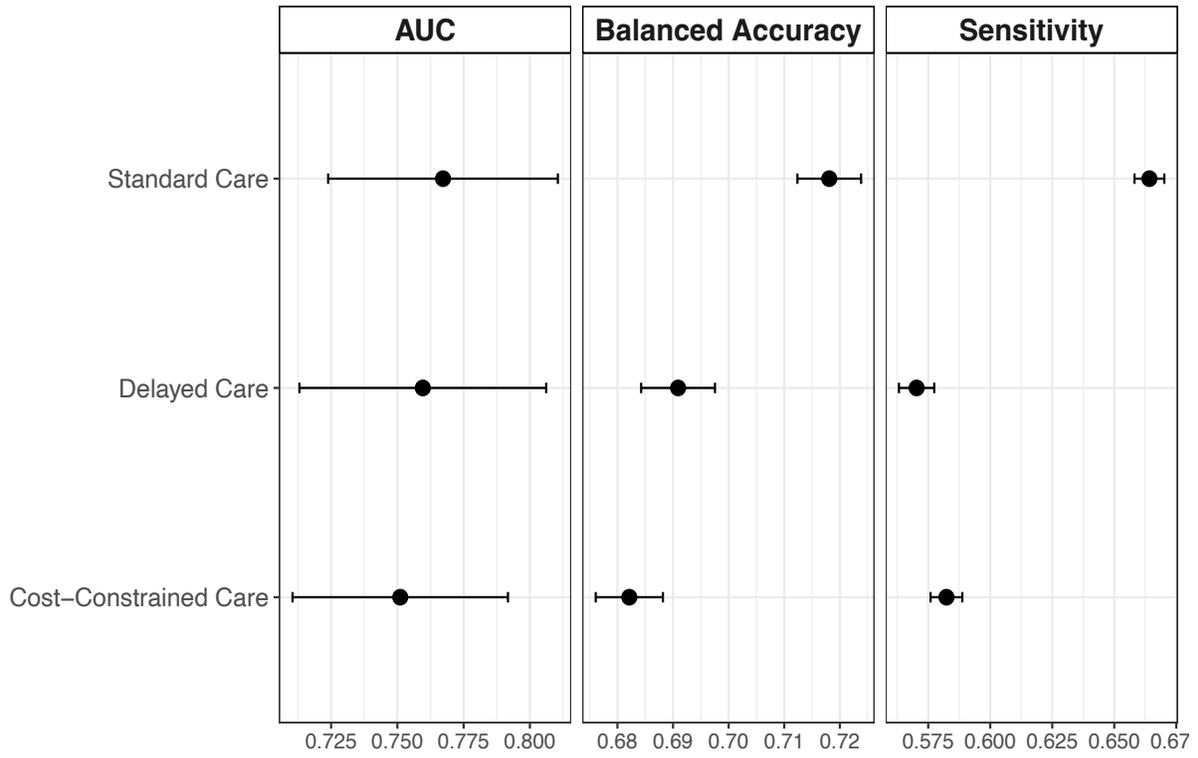

**Figure 4: Mitigation Results for Diabetes Task**

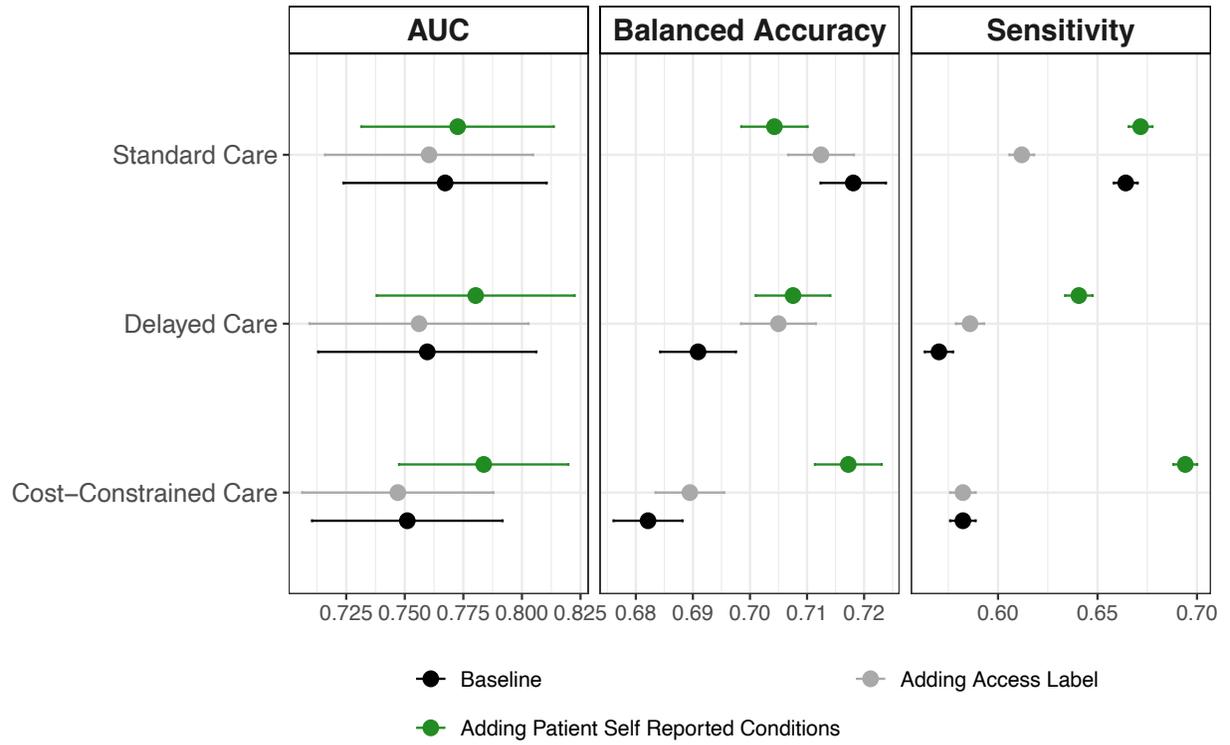

*Supplementary Material*

**Table S1: Percent of Participants Assigned to Access Groups, by Race and Ethnicity Group**

| Access Group | All | Black | Hispanic | White |
|---|---|---|---|---|
| Standard Care | 46.4% | 39.4% | 32.5% | 49.8% |
| Delayed Care | 34.6% | 41.1% | 45.2% | 31.8% |
| Cost-Constrained Care | 42.9% | 49.8% | 56.1% | 39.6% |

**Table S2. Age-Adjusted Self-Reported Condition Prevalence Rate**

| Self-Reported Condition | Standard Care (N = 62,454) | Delayed Care (N=39,882) | Cost-Constrained Care (N=39,882) |
|---|---|---|---|
| Acid Reflux | 26.0 | 34.0 | 36.1 |
| Hypercholesterolemia | 28.7 | 30.7 | 31.4 |
| Astigmatism | 28.5 | 30.9 | 31.9 |
| Hypertension | 27.2 | 31.5 | 32.5 |
| Depression | 20.2 | 33.5 | 36.3 |
| Anxiety and/or Panic Disorder | 13.9 | 24.7 | 26.6 |
| Cataracts | 18.1 | 19.6 | 20.0 |
| Osteoarthritis | 16.6 | 20.0 | 21.1 |
| Asthma | 14.1 | 20.1 | 21.8 |
| Anemia | 14.2 | 20.5 | 22.1 |
| Hypothyroidism | 10.5 | 12.6 | 13.4 |
| Type 2 Diabetes | 7.7 | 11.4 | 12.5 |
| Pre-Diabetes | 7.5 | 10.3 | 10.5 |
| Endometriosis | 4.1 | 6.4 | 7.1 |
| Breast Cancer | 4.4 | 3.9 | 3.7 |

We reported age-adjusted prevalence rates for the top ten most common self-reported conditions. For age adjustment, we find the prevalence rate of each self-reported condition by 10-year age intervals (20, 30, 40, 50, 60, 70, 80, 90, 100+). Then we find the weights for each age group based on the sample age distribution. The final age-adjusted prevalence is the sum of each age interval by its relevant weight. All differences are statistically significant based on a two-sided t-test.

**Table S3: Defining Conditions Using EHR Lab and Measurement Values**

| Condition | Lab/Measurement (Range) |
|---|---|
| Obesity | BMI (>30) |
| Fever | Temperature (>100.4 Fahrenheit or >38 Celsius) |
| Hypocalcemia | Calcium (< 8.5) |
| Hypercalcemia | Calcium (>10.5) |
| Creatinine | Creatinine (>1.3 and <3.4) |
| High Blood Pressure | Systolic BP (>130) or Diastolic BP (>80) |
| Diabetes | Glucose (>126) |
| Tachycardia | Heart Rate (>100) |
| Anemia | Hemoglobin (<12) |
| High Hemoglobin | Hemoglobin (>15) |
| Hypoxemia | Oxygen Value (<.90) |
| Hyperkalemia | Potassium (>6) |
| Hypokalemia | Potassium (<2.5) |
| Tachypneic | Respiratory Rate (>20) |
| Bradypnea | Respiratory Rate (<12) |
| Hypernatremia | Sodium (>145) |
| Hyponatremia | Sodium (<135) |
| Low Urea | Urea (<8) |
| High Urea | Urea (>24) |

**Table S4: Diabetes Sample Characteristics**

|  | Overall (N=52,046) | Standard Care (N=23,705) | Delayed Care (N=18,626) | Cost-Constrained Care (N=22,822) |
|---|---|---|---|---|
| **Access Measures (%)** | | | | |
|   Affordability | 43.8 | 0 | 70.4 | 100.0 |
|   Delayed Care | 35.8 | 0 | 100.0 | 57.4 |
| **(Average) Age** | 53.3 | 58.3 | 45.4 | 49.9 |
| **Gender (%)** | | | | |
|   Male | 30.3 | 36.4 | 22.7 | 25.0 |
|   Female | 66.9 | 61.2 | 73.8 | 71.8 |
|   Other | 2.8 | 2.4 | 3.5 | 3.2 |
| **Race (%)** | | | | |
|   White | 76.0 | 80.9 | 70.2 | 71.1 |
|   Black | 7.7 | 6.0 | 9.7 | 9.5 |
|   Other | 16.2 | 13.0 | 20.0 | 19.3 |
| **Ethnicity (%)** | | | | |
|   Hispanic or Latino | 8.9 | 5.7 | 12.3 | 12.0 |
| **Outcome (%)** | | | | |
|   2-Year Diabetes Incidence | 0.6 | 0.6 | 0.7 | 0.7 |

**Figure S1: Share of Participants who Responded Yes to Access-Specific Questions**

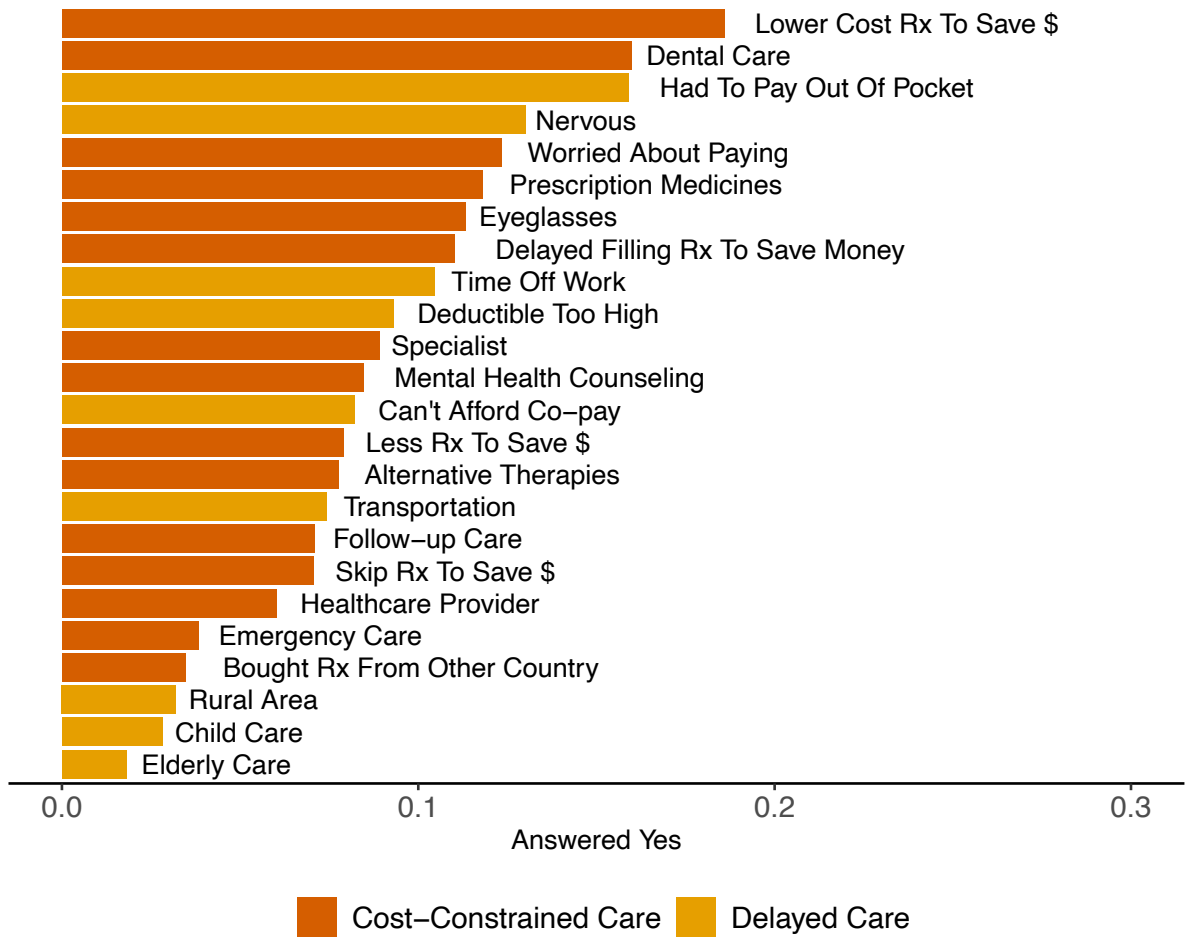

We define participants as having low access if they indicated any affordability issues and had delayed care for the listed reasons. Questions were reported in the All Of Us Health Care Access & Utilization survey.

**Figure S2. Comparison of Reliability of Diagnoses According to EHR, Self-Report, or Both**

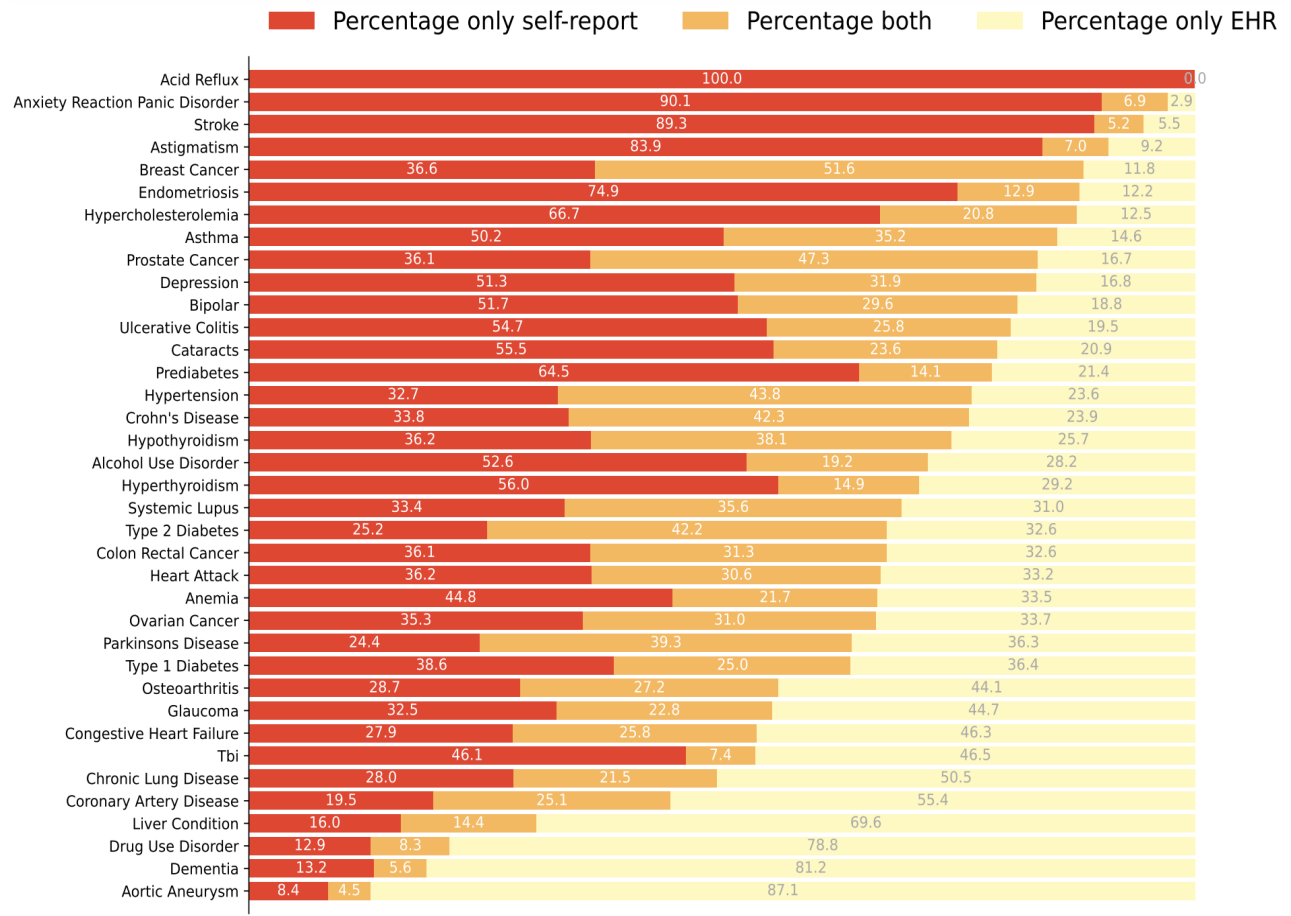

**Table S5: Defining Conditions in the EHR and Survey**

| Condition | EHR Source - SNOMED Codes | Survey Source – Question |
|---|---|---|
| Acid Reflux | 698065002 | Including yourself, who in your family has had acid reflux? |
| Alcohol Use Disorder | 15167005, 66590003, 191812006, 191882002, 191811004, 191813001, 191883007, 191884001, 284591009, 268645007, 7200002, 288031000119105, 713862009, 714829008 | Including yourself, who in your family has had alcohol use disorder? |
| Anemia | 127040003, 191170009, 55995005, 417425009, 38911009, 416180004, 40108008, 398623004, 306058006, 234376007, 111407006, 417048006, 41841004, 84027009, 49472006, 413533008, 199246003, 371315009, 86986002, 191222008, 267513007, 45828008, 267530009, 61261009, 53165003, 87522002, 234347009, 191216004, 271737000, 4854004, 191265009, 413603009, 199247007, 416826005, 49708008, 85746008, 109996008, 109998009, 10619002, 19442009, 199248002, 191217008, 191255003, 191256002, 191142007, 191187006, 189509003, 27342004, 191128004, 191268006, 191261000, 300980002, 234349007, 234348004, 267518003, 267527002, 309742004, 267517008, 310647000, 398937006, 415284008, 413532003, 415283002, 79035003, 44452003, 51071000, 80126007, 313291009, 81711008, 3978000, 59106005, 60504009, 44910003, 44666001, 46737006, 62403005, 65959000, 66612000, 68913001, 78129009, 83414005, 428383000, 328301000119102, 10762261000119105, 713910008, 691421000119108, 724557008, 732960002, 724556004, 789660001, 10741751000119100, 707323002, 707324008 | Including yourself, who in your family has had anemia? |
| Anxiety Reaction or Panic Disorder | 371631005, 35607004, 43150009, 56576003, 82494000, 61212007 | Including yourself, who in your family has had anxiety reaction/panic disorder? |
| Aortic Aneurysm | 14336007, 233985008, 73067008, 67362008, 12232008, 75878002, 233984007, 74883004, 195258006, 433068007, 450781000124101, 16972009, | Including yourself, who in your family has had aortic aneurysm? |

| | 425963007, 426948001, 54160000, 444569004, 2831000119107, 712578006 | |
|---|---|---|
| Asthma | 12428000, 195949008, 281239006, 37981002, 424643009, 409663006, 195967001, 31387002, 5281000124103, 901000119100, 233678006, 225057002, 233683003, 304527002, 425969006, 427603009, 427295004, 426979002, 426656000, 266361008, 427679007, 370221004, 370218001, 370219009, 389145006, 55570000, 405944004, 407674008, 63088003, 423889005, 782520007, 10675391000119101, 10675751000119107, 10674991000119104, 445427006, 72301000119103, 734905008, 735588005, 1751000119100, 703954005, 707444001, 707445000, 707447008, 707511009, 707512002, 707513007, 707979007, 707980005, 707981009, 708090002, 708093000, 708095007, 708096008, 708038006, 707446004, 708094006, 10675431000119106, 10675871000119106, 10676391000119108, 10692681000119108, 10692721000119102, 124991000119109, 125021000119107, 135181000119109, 1741000119102, 125001000119103, 10676431000119103, 135171000119106, 2360001000004109, 10692761000119107 | Including yourself, who in your family has had asthma? |
| Astigmatism | 47099006, 82649003, 68905002, 419774004, 332691000119101, 334661000119104, 338301000119100, 340261000119105, 344071000119102, 345621000119109, 347131000119108, 336341000119104, 341951000119105, 449724005, 449734001 | Including yourself, who in your family has had astigmatism |
| Bipolar | 9340000, 371596008, 35481005, 191621009, 22121000, 192362008, 5703000, 30935000, 59617007, 13746004, 74686005, 191630001, 36583000, 10981006, 191641004, 191639000, 191638008, 191636007, 191634005, 191629006, 191627008, 191625000, 191623007, 40926005, 28884001, 191618007, 191620005, 49468007, 61403008, 63249007, 162004, 16506000, 1196001, 111485001, 13313007, 16295005, 28663008, 34315001, 31446002, 371600003, 371604007, 48937005, 49512000, 53607008, 71984005, 41836007, 38368003, 35846004, 66631006, 68569003, 82998009, 83225003, 85248005, 71294008, 765176007, 767631007, 767632000, 767633005, 767635003, 767636002, 723903001, 723905008, 789061003, 261000119107, 23741000119105 | Including yourself, who in your family has had bipolar disorder? |

| Breast Cancer | 109886000, 93796005, 93745008, 94117003, 372092003, 93876006, 1080101000119105, 1080181000119102, 353511000119101, 865954003, 109887009, 278052009, 278054005, 188050009, 188147009, 188154003, 188155002, 188156001, 188151006, 188152004, 188157005, 188153009, 254840009, 254837009, 286894008, 254838004, 286895009, 427685000, 372094002, 372064008, 372095001, 372096000, 373083005, 372137005, 373082000, 373088001, 373090000, 314955001, 417181009, 408643008, 93215006, 403946000, 353431000119107, 459391000124109, 1080241000119104, 1080261000119100, 1081551000119106, 713609000, 718220008, 444712000, 447782002, 448952004, 706970001, 708921005 | Including yourself, who in your family has had breast cancer? |
|---|---|---|
| Coronary Artery Disease | 87343002, 50570003, 53741008, 28574005, 429673002, 451041000124103, 451361000124102, 213037002, 123641001, 251024009, 194821006, 194842008, 233817007, 23687008, 371803003, 371804009, 371805005, 429245005, 414024009, 67682002, 92517006, 408546009, 75398000, 117051000119103, 719678003, 139011000119104, 724431008, 732230001, 443502000, 442224005, 442421004, 445512009, 473443007, 285151000119108 | Including yourself, who in your family has had coronary artery/coronary heart disease? |
| Cataracts | 1412008, 53889007, 8656007, 38583007, 111515007, 193570009, 52421005, 11422002, 34361001, 193600001, 43959009, 399305009, 849000, 204127006, 79410001, 43972005, 39450006, 76562003, 193602009, 78875003, 193615000, 34533008, 5318001, 193589009, 65720003, 51044000, 28550007, 10999321000119108, 10999361000119103, 15737041000119101, 15737081000119106, 15737121000119108, 15737321000119105, 15737361000119100, 15737921000119106, 15738481000119102, 15738521000119102, 678601000119100, 678621000119109, 678671000119105, 678681000119108, 10999281000119103, 678611000119102, 678691000119106, 680531000119103, 204128001, 193576003, 193577007, 193590000, 193609000, 264443002, 315353005, 39944005, 420756003, 421920002, 359766000, 410568009, 77873007, 95722004, 231970007, 766834007, 331781000119101, 335861000119104, 337401000119107, | Including yourself, who in your family has had cataracts? |
| | 341471000119109, 346721000119108, 347471000119101, 347531000119100, 347561000119108, 347591000119101, 1078791000119109, 1078801000119105, 15639401000119105, 15639441000119107, 15639481000119102, 718851007, 15737161000119103, 15737201000119108, 15737241000119105, 15737281000119100, 346691000119104, 346701000119104, 347481000119103, 347541000119109, 347601000119108, 678581000119109, 678591000119107, 16477021000119103, 15990111000119104, 15990031000119108, 816120008, 816119002, 441622000, 445213003, 446474007, 699521008, 342291000119102 | |
| Hyper-cholesterolemia | 267432004, 238079002, 13644009, 238076009, 238080004, 398036000 | Including yourself, who in your family has had high cholesterol? |
| Colon Cancer | 93984006, 109838007, 93761005, 363412000, 94006002, 109839004, 93826009, 363410008, 363409003, 363408006, 363351006, 187757001, 187760008, 254582000, 254586002, 269533000, 363406005, 363407001, 363414004, 363413005, 312112002, 315058005, 314965007, 781382000, 681601000119101, 721695008, 713573006, 726654006 | Including yourself, who in your family has had colon cancer/rectal cancer? |
| Crohn's Disease | 7620006, 71833008, 56689002, 34000006, 196977009, 426549001, 50440006, 414154002, 3815005, 38106008, 397173003, 1085801000119106 | Including yourself, who in your family has had Crohn's disease? |
| Dementia | 40425004, 191519005, 191449005, 191451009, 70936005, 191455000, 51928006, 26929004, 281004, 278857002, 429998004, 14070001, 428051000124108, 230287006, 52448006, 312991009, 416780008, 416975007, 6475002, 66108005, 288631000119104, 16276361000119109, 442344002, 79341000119107, 1581000119101, 1591000119103, 105421000119105 | Including yourself, who in your family has had dementia (includes Alzheimer's, vascular, etc.)? |
| Depression | 268621008, 191611001, 78667006, 68019004, 191613003, 36474008, 191630001, 430852001, 191659001, 191610000, 191634005, 191629006, 191627008, 191604000, 35489007, 76441001, 14070001, 19527009, 162722001, 15639000, 18818009, 247803002, 2618002, 191616006, 191495003, 192080009, 300706003, 33078009, 33135002, 33736005, 30605009, 310495003, | Including yourself, who in your family has had depression? |
| | 310497006, 310496002, 370143000, 28475009, 320751009, 42810003, 79298009, 319768000, 84788008, 85080004, 84760002, 40379007, 58703003, 596004, 73867007, 46244001, 63412003, 36923009, 66344007, 832007, 83176005, 87414006, 70747007, 71336009, 75084000, 87512008, 231504006, 16264821000119108, 16264901000119109, 16265301000119106, 765176007, 719592004, 719593009, 712823008, 94631000119100, 726772006, 442057004, 251000119105, 450714000, 281000119103 | |
| Drug Use Disorder | 191821007, 191833002, 191837001, 191875001, 191873008, 15167005, 191831000, 191827006, 191912005, 191893000, 191894006, 191895007, 66590003, 191812006, 191882002, 191936009, 191871005, 31956009, 191811004, 26416006, 89765005, 21647008, 75544000, 191813001, 191934007, 231461004, 191918009, 191816009, 191820008, 85005007, 191883007, 191832007, 191819002, 191937000, 191839003, 191901005, 191884001, 268640002, 427327003, 836439001, 2403008, 280982009, 191865004, 191887008, 191869005, 191867007, 191889006, 191924003, 191938005, 284591009, 426873000, 425533007, 268645007, 5002000, 428219007, 51339003, 56294008, 7200002, 6525002, 231478008, 231477003, 231458000, 231473004, 288031000119105, 713862009, 714829008, 16077091000119107, 724688003, 724698009, 441527004, 442406005, 1081000119105, 153501000119105, 1461000119109, 1471000119103, 153491000119103 | Including yourself, who in your family has had a drug use disorder? |
| Endometriosis | 53913001, 22611009, 5562006, 198251001, 266589005, 76376003, 129103003, 237116001, 17829005, 26681001, 50993001, 52533003 | Including yourself, who in your family has had endometriosis? |
| Glaucoma | 33647009, 29538005, 193548006, 71111008, 34623005, 84333006, 77075001, 392288006, 302895007, 30041005, 37155002, 66747002, 45623002, 46168003, 65460003, 111514006, 23986001, 35472004, 1207009, 68241007, 193561006, 84494001, 50485007, 12239301000119102, 12239421000119101, 15633321000119108, 15640441000119104, 15679761000119102, 15736441000119108, 15739441000119101, 12239501000119106, 12239461000119106, 15633281000119103, 678471000119107, 833278008, | Including yourself, who in your family has had glaucoma? |
|  | 1654001, 21571006, 21928008, 193531003, 193533000, 314784002, 392291006, 392029006, 314033007, 392352004, 92829008, 93435005, 95250000, 95717004, 232080006, 232086000, 334321000119101, 338721000119108, 339921000119104, 341641000119100, 344491000119107, 345291000119109, 346861000119107, 718851007, 15736721000119106, 787051000, 787052007, 332871000119103, 338481000119100, 344251000119103, 678411000119104, 788946007, 204113001, 444863008 |  |
| Heart Attack | 57054005, 1755008, 54329005, 58612006, 73795002, 70211005, 65547006, 76593002, 70422006, 233840006, 307140009, 314207007, 59063002, 401314000, 401303003, 79009004, 22298006, 16837681000119104, 285981000119103, 17531000119105, 703211006, 703360004 | Including yourself, who in your family has had a heart attack? |
| Heart failure | 5148006, 82523003, 42343007, 194779001, 194767001, 7401000175100, 10633002, 426263006, 426611007, 88805009, 92506005, 66989003, 15629541000119106, 67441000119101, 96311000119109, 698296002, 23341000119109, 15781000119107, 698594003 | Including yourself, who in your family has had congestive heart failure? |
| Hypertension | 198985009, 198999008, 199002002, 198967002, 198965005, 198966006, 1201005, 41114007, 198946002, 194785008, 38341003, 123799005, 78975002, 31992008, 59621000, 199005000, 198941007, 198942000, 46764007, 198983002, 198984008, 398254007, 237279007, 28119000, 10725009, 23130000, 198986005, 194791005, 194783001, 194788005, 288250001, 307632004, 371125006, 48146000, 48194001, 429457004, 52698002, 56218007, 57684003, 72022006, 8762007, 73410007, 63287004, 65518004, 367390009, 67359005, 70272006, 37618003, 78808002, 86041002, 95605009, 765182005, 16229371000119106, 461301000124109, 443482000, 82771000119102, 541000119105, 132721000119104, 697929007, 697930002, 40521000119100, 706882009 | Including yourself, who in your family has had high blood pressure (hypertension)? |
| Hyperthyroid | 29028009, 89719007, 60216004, 57777000, 26389007, 69329005, 90739004, 267374005, 190255006, 73869005, 237498007, 190247004, 190241003, 27538003, 286909009, 237508001, 237510004, 34486009, 427970008, 55807009, 5604000, 353295004, 62052002, 87232008, 137421000119106 | Including yourself, who in your family has had hyperthyroidism? |
| Hypothyroid | 237527007, 190268003, 88273006, 27059002, 111566002, 190279008, 40930008, 237528002, 237519003, 237515009, 237520009, 237567008, 26692000, 428165003, 50375007, 43153006, 54823002, 39444001, 82598004, 83986005, 89261000, 40539002, 4641009, 367631000119105, 57185003 | Including yourself, who in your family has had hypothyroidism? |
| Liver Condition | 235869004, 235866006, 30188007, 67656006, 1761006, 50325005, 39400004, 34798003, 111896003, 17890003, 61977001, 266468003, 15230009, 235856003, 128241005, 262802005, 70668000, 51292008, 420054005, 82385007, 235867002, 50711007, 72925005, 80515008, 76795007, 111891008, 94381002, 128302006, 17920008, 41889008, 34736002, 33167004, 408335007, 76783007, 197268000, 9953008, 95214007, 41309000, 79720007, 27916005, 186626002, 186639003, 199118005, 431674004, 13891000, 89166001, 435101000124104, 838311008, 109820009, 109841003, 109842005, 111331000, 102626001, 10295004, 197270009, 197284004, 197315008, 197356006, 13923006, 109559009, 12368000, 197271008, 197293003, 197352008, 197354009, 197355005, 197279005, 197321007, 197358007, 197359004, 199203001, 19943007, 186698009, 187777008, 187767006, 262796008, 262799001, 25102003, 300332007, 300331000, 300338006, 302919001, 126851005, 278527001, 31712002, 33688009, 427399008, 371067004, 371139006, 48036004, 4637005, 432908002, 53476006, 57412004, 328383001, 40468003, 85057007, 88518009, 89580002, 58008004, 92186001, 37871000, 59927004, 93870000, 93469006, 408646000, 61860000, 62484002, 63246000, 65617004, 36760000, 66071002, 67251005, 3738000, 76301009, 76281005, 77981007, 38739001, 424263008, 86514004, 428187007, 75183008, 431222008, 235875008, 235881000, 235886005, 235896001, 235899008, 235876009, 235888006, 235910007, 235859005, 235878005, 235911006, 235912004, 235903001, 236004002, 243978007, 765482002, 714253009, 715140008, 723360007, 716203000, 717232005, 442191002, 442685003, 447109003, 1691000119104, 831000119103, 707341005, 708198006, 708248004 | Including yourself, who in your family has had a liver condition (e.g., cirrhosis)? |
| Chronic Long Disease | 13645005, 63480004, 195949008, 195951007, 74417001, 185086009, 57686001, 87433001, 61937009, 33325001, 77690003, 195957006, | Including yourself, who in your family has had |

| | 196001008, 195953005, 425748003, 47938003, 4981000, 313297008, 313296004, 313299006, 89099002, 68328006, 785736001, 1751000119100, 106001000119101, 293241000119100, 10692761000119107 | chronic lung disease (COPD, emphysema, or bronchitis)? |
|---|---|---|
| Systemic Lupus | 55464009, 196138005, 201436003, 309762007, 52042003, 54072008, 36402006, 68815009, 403486000, 76521009, 77753005, 95644001, 239887007, 239888002 | Including yourself, who in your family has had systemic lupus? |
| Osteoarthritis | 67315001, 20075001, 90860001, 201829007, 75468006, 428768008, 201831003, 60937000, 429420005, 34427002, 201834006, 201819000, 33262002, 26241001, 201645006, 201849003, 33420007, 22193007, 201837004, 10948005, 201852006, 396275006, 201855008, 18834007, 201847001, 48210000, 77547008, 36071006, 68859000, 267970006, 67437007, 69195002, 201646007, 77994009, 91240008, 429419004, 201625003, 63198006, 201634008, 68675004, 26538006, 1074371000119104, 1074381000119101, 1074421000119105, 1074431000119108, 1074441000119104, 12241631000119102, 12241671000119104, 14380001000004102, 15720481000119102, 323251000119103, 323261000119101, 109668000, 111242007, 239832006, 239833001, 239862000, 239865003, 239867006, 239877008, 123798002, 202688001, 202693003, 239872002, 239873007, 239828000, 239863005, 239866002, 239874001, 190828008, 190842000, 28736004, 201835007, 201858005, 201826000, 201857000, 33952002, 267889007, 268054009, 267890003, 373623009, 275324008, 50921008, 43132002, 43829003, 313257005, 53332000, 309246000, 72275000, 82300000, 8847002, 37895003, 73877009, 36540006, 387802007, 78675000, 387800004, 239868001, 239878003, 239881008, 239869009, 239870005, 239880009, 112961000119103, 112971000119109, 112981000119107, 112991000119105, 113001000119106, 113011000119109, 113021000119102, 113041000119108, 321341000119107, 1075021000119104, 16039191000119102, 16039391000119100, 1074461000119100, 1074491000119107, 303041000119104, 318661000119102, 318671000119108, 318681000119106, | Including yourself, who in your family has had osteoporosis? |
| | 318691000119109, 318711000119107, 318721000119100, 318731000119102, 323291000119108, 323311000119107, 323321000119100, 15749161000119108, 323301000119109, 318641000119101, 318701000119109, 713826008, 1074111000119109, 1074161000119107, 1074521000119109, 12367411000119102, 12367461000119104, 15637471000119109, 15722441000119106, 15722561000119106, 16008111000119108, 16008191000119104, 16016631000119106, 16583281000119101, 16583321000119106, 16583361000119101, 724604003, 12367361000119109, 1074481000119109, 1074511000119102, 16039151000119107, 789000004, 442925003, 442928001, 442942009, 443001001, 443002008, 443524000, 445478004, 442884002, 1074121000119102, 1074131000119104, 1074141000119108, 1074151000119105, 1074531000119107, 16010951000119105, 16039431000119105, 16039631000119108, 303861000119106, 318651000119104, 735602002, 735607008, 737057000, 735603007, 1074071000119102, 735605000, 450521003, 699262001, 700293006, 703051004, 304321000119105 | |
| Ovarian Cancer | 94455000, 93934004, 254850005, 254851009, 254852002, 254863004, 254849005, 363443007, 369523007, 369530001, 763131005, 722684000, 827162007, 718220008 | Including yourself, who in your family has had polycystic ovarian syndrome? |
| Parkinson's Disease | 49049000, 718685006, 715345007 | Including yourself, who in your family has had Parkinson's disease? |
| Prediabetes | 714628002 | Including yourself, who in your family has had prediabetes? |
| Prostate Cancer | 93974005, 254900004, 427492003, 399490008, 399068003, 459381000124106, 1098981000119101, 722103009, 712849003 | Including yourself, who in your family has had prostate cancer? |
| Stroke | 70936005, 230690007, 14070001, 16002151000119107, 16002231000119106, | Including yourself, who in your |
| | 16002271000119109, 16002351000119105, 16002391000119100, 16002431000119105, 16002511000119104, 329621000119105, 230691006, 230715005, 195217004, 195212005, 195213000, 195216008, 371041009, 371040005, 413758000, 57981008, 422504002, 457551000124104, 16371781000119100, 329671000119106, 16026951000119102, 716051003, 16000511000119103, 16002111000119106, 292671000119104, 292691000119103, 329641000119104, 329651000119102, 330791000119108, 16024111000119109, 9901000119100, 99451000119105 | family has had a stroke? |
| TBI | 62564004, 269144002, 62106007, 53267002, 127310009, 127309004, 127308007, 870548008, 870549000, 870550000, 870551001, 110030002, 210038008, 209827006, 23713006, 230763008, 262689001, 262688009, 127298000, 127299008, 127295002, 127300000, 127301001, 34663006, 9015001, 329671000119106, 447396006, 698620007, 698621006 | Including yourself, who in your family has had traumatic brain injury (TBI)? |
| Type 1 Diabetes | 46635009, 313435000, 23045005, 190368000, 28032008, 426875007, 609564002, 609566000, 31321000119102 | Including yourself, who in your family has had type 1 diabetes? |
| Type 2 Diabetes | 44054006, 190389009, 237627000, 237599002, 313436004, 359642000, 81531005, 609567009, 1481000119100 | Including yourself, who in your family has had type 2 diabetes? |
| Ulcerative Colitis | 52506002, 13470001, 64766004, 15342002, 128600008, 78324009, 444546002, 444548001, 441971007, 442159003, 445243001 | Including yourself, who in your family has had ulcerative colitis? |